\documentclass[useAMS,usenatbib]{mn2e}

\usepackage{amssymb,graphicx,multirow,txfonts}

\newcommand{\etal}{et~al.}
\newcommand{\PVdblt}{{\rm P}\kern 0.1em{\sc v}~$\lambda\lambda 1117, 1128$}
\newcommand{\CaIIdblt}{{\rm Ca}\kern 0.1em{\sc ii}~$\lambda\lambda 3934, 3969$}
\newcommand{\AlIIIdblt}{{\rm Al}\kern 0.1em{\sc iv}~$\lambda\lambda 1855, 1863$}
\newcommand{\CIVdblt}{{\rm C}\kern 0.1em{\sc iv}~$\lambda\lambda 1548, 1550$}
\newcommand{\MgIIdblt}{{\rm Mg}\kern 0.1em{\sc ii}~$\lambda\lambda 2796, 2803$}
\newcommand{\NVdblt}{{\rm N}\kern 0.1em{\sc v}~$\lambda\lambda 1238, 1242$}  
\newcommand{\SVIdblt}{{\rm S}\kern 0.1em{\sc vi}~$\lambda\lambda 933, 944$} 
\newcommand{\OVIdblt}{{\rm O}\kern 0.1em{\sc vi}~$\lambda\lambda 1031, 1037$} 
\newcommand{\SiIIdblt}{{\rm Si}\kern 0.1em{\sc ii}~$\lambda\lambda 1190, 1193$} 
\newcommand{\SiIVdblt}{{\rm Si}\kern 0.1em{\sc iv}~$\lambda\lambda 1393, 1402$} 
\newcommand{\AlI}{\hbox{{\rm Al}\kern 0.1em{\sc i}}}
\newcommand{\AlII}{\hbox{{\rm Al}\kern 0.1em{\sc ii}}}
\newcommand{\AlIII}{{\hbox{\rm Al}\kern 0.1em{\sc iii}}}
\newcommand{\CaII}{\hbox{{\rm Ca}\kern 0.1em{\sc ii}}}
\newcommand{\CII}{\hbox{{\rm C}\kern 0.1em{\sc ii}}}
\newcommand{\CIIe}{\hbox{{\rm C$^{\ast}$}\kern 0.1em{\sc ii}}}
\newcommand{\CIII}{\hbox{{\rm C}\kern 0.1em{\sc iii}}}
\newcommand{\CIV}{\hbox{{\rm C}\kern 0.1em{\sc iv}}}
\newcommand{\CV}{\hbox{{\rm C}\kern 0.1em{\sc v}}}
\newcommand{\HI}{\hbox{{\rm H}\kern 0.1em{\sc i}}}
\newcommand{\HII}{\hbox{{\rm H}\kern 0.1em{\sc ii}}}
\newcommand{\Lya}{\hbox{{\rm Ly}\kern 0.1em$\alpha$}}
\newcommand{\Lyb}{\hbox{{\rm Ly}\kern 0.1em$\beta$}}
\newcommand{\Lyg}{\hbox{{\rm Ly}\kern 0.1em$\gamma$}}
\newcommand{\Lyd}{\hbox{{\rm Ly}\kern 0.1em$\delta$}}
\newcommand{\HeI}{\hbox{{\rm He}\kern 0.1em{\sc i}}}
\newcommand{\HeII}{\hbox{{\rm He}\kern 0.1em{\sc ii}}}
\newcommand{\FeI}{\hbox{{\rm Fe}\kern 0.1em{\sc i}}}
\newcommand{\FeII}{\hbox{{\rm Fe}\kern 0.1em{\sc ii}}}
\newcommand{\FeIII}{\hbox{{\rm Fe}\kern 0.1em{\sc iii}}}
\newcommand{\MnII}{\hbox{{\rm Mn}\kern 0.1em{\sc ii}}}
\newcommand{\MgI}{\hbox{{\rm Mg}\kern 0.1em{\sc i}}}
\newcommand{\MgII}{\hbox{{\rm Mg}\kern 0.1em{\sc ii}}}
\newcommand{\MgIII}{\hbox{{\rm Mg}\kern 0.1em{\sc iii}}}
\newcommand{\NI}{\hbox{{\rm N}\kern 0.1em{\sc i}}}
\newcommand{\NII}{\hbox{{\rm N}\kern 0.1em{\sc ii}}}
\newcommand{\NIII}{\hbox{{\rm N}\kern 0.1em{\sc iii}}}
\newcommand{\NV}{\hbox{{\rm N}\kern 0.1em{\sc v}}}
\newcommand{\OVI}{\hbox{{\rm O}\kern 0.1em{\sc vi}}}
\newcommand{\OI}{\hbox{{\rm O}\kern 0.1em{\sc i}}}
\newcommand{\OII}{\hbox{[{\rm O}\kern 0.1em{\sc ii}]}}
\newcommand{\OIII}{\hbox{[{\rm O}\kern 0.1em{\sc iii}]}}
\newcommand{\OIV}{\hbox{{\rm O}\kern 0.1em{\sc iv}]}}
\newcommand{\SI}{{\rm S}\kern 0.1em{\sc i}}
\newcommand{\SIV}{{\rm S}\kern 0.1em{\sc iv}}
\newcommand{\SVI}{{\rm S}\kern 0.1em{\sc vi}}
\newcommand{\SiI}{\hbox{{\rm Si}\kern 0.1em{\sc i}}}
\newcommand{\SiII}{\hbox{{\rm Si}\kern 0.1em{\sc ii}}}
\newcommand{\SiIII}{\hbox{{\rm Si}\kern 0.1em{\sc iii}}}
\newcommand{\SiIV}{\hbox{{\rm Si}\kern 0.1em{\sc iv}}}
\newcommand{\SII}{\hbox{{\rm S}\kern 0.1em{\sc ii}}}
\newcommand{\SIII}{\hbox{{\rm S}\kern 0.1em{\sc iii}}}
\newcommand{\NaI}{\hbox{{\rm Na}\kern 0.1em{\sc i}}}
\newcommand{\TiII}{\hbox{{\rm Ti}\kern 0.1em{\sc ii}}}
\newcommand{\ZnII}{\hbox{{\rm Zn}\kern 0.1em{\sc ii}}}
\newcommand{\CrII}{\hbox{{\rm Cr}\kern 0.1em{\sc ii}}}
\newcommand{\kms}{\hbox{km~s$^{-1}$}}
\newcommand{\cmsq}{\hbox{cm$^{-2}$}}

\title[Galaxy group associated with DLA]{Galaxy group at \mbox{$z$}=0.3 associated with the damped Lyman alpha system towards quasar Q1127--145}
\author[G. G. Kacprzak et al.]{Glenn G. Kacprzak,$^{1}$\thanks{gkacprzak@astro.swin.edu.au} 
Michael T. Murphy,$^{1}$ and 
Christopher W. Churchill$^{2}$\\
$^{1}$ Centre for Astrophysics and Supercomputing, Swinburne University of Technology, PO Box 218, Victoria 3122, Australia\\
$^{2}$ Department of Astronomy, New Mexico State University, Las Cruces, NM 88003}
\begin{document}

\date{ December 14 2009}

\pagerange{\pageref{firstpage}--\pageref{lastpage}} \pubyear{2010}

\maketitle

\label{firstpage}

\begin{abstract}

We performed a spectroscopic galaxy survey, complete to
$m_{F814W}$$\leq$20.3 ($L_B>0.15L_B^{\star}$ at $z=0.3$), within
100$\times$100$''$ of the quasar Q1127--145 ($z_{em}=1.18$). The
VLT/UVES quasar spectrum contains three $z_{abs}$$<$0.33 {\MgII}
absorption systems. We obtained eight new galaxy redshifts, adding to
the four previously known, and galaxy star formation rates (SFRs) and
metallicities were computed where possible. A strong {\MgII} system
[$W_r(2796)=1.8$~\AA], which is a known damped Ly$\alpha$ absorber
(DLA), had three previously identified galaxies; we found two
additional galaxies associated with this system. These five galaxies
form a group with diverse properties, such as a luminosity range of
$0.04\leq L_B\leq0.63 L_B^{\star}$, an impact parameter range of
$17\leq D \leq 241$~kpc and velocity dispersion of
$\sigma=115$~\kms. The DLA group galaxy redshifts span beyond the
350~{\kms} velocity spread of the metallic absorption lines of the DLA
itself. The two brightest group galaxies have SFRs of $\sim$few
$M_{\odot}$~yr$^{-1}$ and should not have strong winds. We have
sufficient spectroscopic information to directly compare three of the
five group galaxies' (emission-line) metallicities with the DLA
(absorption) metallicity: the DLA metallicity is 1/10th solar,
substantially lower than the three galaxies' which range between less
than 1/2 solar to solar metallicity. HST/WFPC--2 imaging shows
perturbed morphologies for the three brightest group galaxies, with
tidal tails extending $\sim$25 kpc. We favor a scenario where the DLA
absorption originates from tidal debris in the group environment.

Another absorber exhibits weak {\MgII} absorption
[$W_r(2796)=0.03$~\AA] and had a previously identified galaxy at a
similar redshift. We have identified a second galaxy associated with
this system. Both galaxies have solar metallicities and unperturbed
morphologies in the HST/WFPC--2 image. The SFR of one galaxy is much
lower than expected for strong outflows. Finally, we have also
identified five galaxies at large impact parameters with no associated
{\MgII} absorption [$W_r(2796) \lesssim 5.7$~m{\AA}, 3~$\sigma$] in
the spectrum of Q1127--145.

\end{abstract}

\begin{keywords}
---galaxies: ISM, haloes, interactions  ---quasars: absorption lines.
\end{keywords}

\section{Introduction}

Absorption lines detected in the spectra of background quasars and
gamma ray bursts remain one of the best probes of intervening
multiphase gas throughout the Universe. Pioneering work of
\citet{bergeron88} and \citet{bb91} led to the first galaxies
identified in close proximity to a quasar sight-line and at the same
redshift as metal-enriched absorption traced by the {\MgIIdblt}
doublet. Since then, there has been numerous studies of {\MgII}
absorption line systems aimed at interpreting the properties of galaxy
halos at a variety of redshifts
\citep[e.g.,][]{lebrun93,sdp94,csv96,archiveII,steidel02,ellison03,bouche06,zibetti07,kacprzak08,chen08,barton09,rubin09a,pollack09,menard09}.

$\hbox{{\rm Mg}\kern 0.1em{\sc ii}}$~absorption~lines are ideal for studying a large dynamic range
of structures and environments in and around galaxies since they trace
low ionization metal-enriched gas with neutral hydrogen column
densities of $10^{16} \lesssim \hbox{N(\HI)} \lesssim
10^{22}$~{\cmsq}\citep{archiveI,weakII}. This large density range
allows for detections of {\MgII} in absorption out to $\sim120$~kpc
from the host galaxy \citep{zibetti07,chen08,kacprzak08}.


Significant theoretical efforts have employed semi-analytical models
and single halo galaxy simulations to interpret and understand
absorption systems
\citep[e.g.,][]{mo96,burkert00,lin00,maller04,chelouche08,chen08,tinker08,kaufmann09}.
These models and simulations have helped constrain halo sizes,
covering fractions, gas kinematics, physical gas conditions,
etc. However, the majority of these studies modeled galaxies as
isolated systems/halos and lack the important dynamic influences of
the cosmic structure and local environments, which also may contribute
to a significant fraction of the detected {\MgII} absorption
\citep{kacprzak10}.


Halo gas masses and cross sections are suggested to increase due to
tidal streams produced by the interactions/minor mergers and/or
increased star formation-induced winds caused by gas rich minor
mergers \citep{york86,rubin09a}. {\MgII} gas outflowing at
$\sim500$~{\kms} has been detected in winds of galaxies at $z\sim1$
\citep[e.g.,][]{tremonti07,weiner09,rubin09b}. Winds have also been
suggested to be responsible for high equivalent width {\MgII}
absorbers \citep{bouche06}.  Evidence of galaxy interactions producing
{\MgII} absorption was discussed by \citet{kacprzak07} who reported a
suggestive correlation between the {\MgII} rest equivalent width,
$W_r(2796)$, and the galaxy morphological asymmetries normalized by
impact parameter. They suggest that perturbations from minor galaxy
mergers may be responsible for producing low equivalent width systems
[$W_r(2796)<1.5$~\AA]. These results are consistent with low redshift
{\HI} surveys where galaxies having a perturbed/warped disk, from a
previous or ongoing minor merger, also have more extended {\HI}
disks/halos
\citep{puche92,swaters97,rand00,fraternali02,chynoweth08,sancisi08}. The
aforementioned results suggest that galaxy environments may play a
role in the metal enrichment of galaxy halos.

\citet{lopez08} performed the first statistical environmental study of
absorption systems associated with 442 x-ray selected galaxy clusters
($z=$0.3--0.9) out to transverse distances of 2~h$^{-1}$~Mpc.  It was
determined that galaxy clusters produce a factor of 15 over-abundance
of strong equivalent width systems [$W_r(2796) > 2$~\AA] compared to
field galaxies. This over-abundance is higher in the centers of
clusters than in the outer parts and also increases with cluster
mass. In contrast, the $dN/dz$ of weak {\MgII} systems [$W_r(2796) <
0.3$~\AA] in clusters is consistent with those derived from
environmentally unbiased samples. \citet{lopez08} argue that the
detected over-abundance of strong systems is a result of the
over-density of galaxies in a cluster region. The lack of an
over-abundance of weak systems may imply they were destroyed by the
cluster environment. \citet{padilla09} modeled these results and found
environmental evidence of truncated {\MgII} halo sizes as a function
of cluster radii. Their models require a median {\MgII} halo size of
$r < 10$~h$^{-1}$~kpc, compared to $35-85$~h$^{-1}$~kpc for field
galaxies, in order to reproduce the observed absorption-line
statistics of \citet{lopez08}. These x-ray selected clusters represent
more extreme environments than in galaxy groups where the majority of
galaxies reside.  In x-ray clusters, ram-pressure stripping is an
important environmental effect since the intracluster medium and
galaxy velocities are much higher than in galaxy groups where
ram-pressure stripping is negligible \citep{mulchaey98}. However, in
groups the galaxy velocities are smaller and the interactions and
mergers more frequent, resulting in increased gas covering fractions
of the cool intragroup gas \citep{zabludoff98}.

A significant fraction of strong {\MgII} absorption systems are damped
Lyman alpha systems (DLA) \citep{rao03}.  Since the discovery of DLAs
\citep{wolfe86} their host galaxy properties and environments have
remained largely unknown. Only a small fraction of DLA hosts have been
identified and appear to be isolated galaxies in close proximity of
the quasar LOS \citep{lacy03,moller02,rao03,chun06}. Models support an
array of origins of the DLA absorbing gas from rotating thick disks
\citep[e.g.,][]{prochaska97,prochaska98,prochaska02}, gas rich dwarf
galaxies \citep{matteucci97}, irregular protogalactic clumps
\citep{haehnelt98}, and tidal gas or processes such as superwinds and
outflows \citep{zwaan08}.

In this paper, we perform a spectroscopic survey of the galaxies in
the Q1127$-$145 quasar field. A VLT/UVES quasar spectrum shows that
there are three absorption systems in this field, one of which is DLA
system which has three previously identified galaxies at a similar
redshift. However, this field contains many unidentified bright
galaxies within 50$''$ of the quasar line of sight. We perform a
spectroscopic survey, to a limiting magnitude of $m_{F814W}\leq 20.3$,
in an attempt to obtain spectroscopic redshifts for the remaining
galaxies within the field.  In \S~\ref{sec:data} we describe our
sample and analysis. In \S~\ref{sec:results} we present the results of
our redshift survey. We discuss morphologies of the galaxies and we
also compute galaxy star formation rates (SFRs) and emission line
metallicities when possible. We compare galaxy metallicities to the
absorption line metallicity derived for the DLA. In \S~\ref{sec:dis},
we discuss the possible origins of the {\MgII} absorption and our
concluding remarks are in \S~\ref{sec:conclusion}. Throughout we adopt
an H$_{\rm 0}=70$~\kms Mpc$^{-1}$, $\Omega_{\rm M}=0.3$,
$\Omega_{\Lambda}=0.7$ cosmology.

\section{Target Field and Observations}
\label{sec:data}

Q1127$-$145 is a bright (V=16.9 mag) $z_{em}=$1.18, gigahertz-peaked
radio source with a $\sim$300~kpc jet seen in the x-ray and in
multi-frequency radio observations
\citep[see][]{siemiginowska02,siemiginowska07}.  A VLT/UVES spectrum
of the quasar contains three {\MgII} absorption systems:
$z_{abs}=$0.190973 (Evans {\etal}, in prep), 0.312710 \citep{bb91} and
0.328266 \citep{narayanan07}. To date, no galaxies have been found to
have similar redshift as the $z_{abs}=$0.190973 {\MgII}
absorption. One galaxy associated with the $z_{abs}=$0.328266
absorption system (labeled in this paper as G5) was recently
spectroscopically confirmed by \citet{kacprzak10}. 

From a {\it HST}/FOS UV spectrum, the $z_{abs}=$0.312710 was determined
to be a DLA with $N_{HI}=5.1\pm0.9\times10^{21}$~cm$^{-2}$
\citep{rao00}.  \citet{bb91} spectroscopically identified two galaxies
(labeled in this paper as G2 and G4) to be at a similar redshift as
the $z_{abs}=$0.312710 absorption system. Since G2 is closer to the
quasar line of sight (LOS) and has more significant star formation
than G4, G2 was favored as the absorbing galaxy.  \citet{lane98} later
obtained a spectroscopic redshift of another galaxy, called G1 here,
which is also consistent with the absorption redshift and was then
favored as the absorbing galaxy due to its closer proximity to the
quasar LOS than G2 and G4.  Additional multi-band imaging studies
claim to have detected low surface brightness emission around the
quasar and also a possible underlying galaxy $0.6''$ from the quasar
LOS \citep{nestor02,rao03,chun06}. However, it is possible that the
low surface brightness signal is coming from radio-loud/x--ray
emitting quasar host galaxy at $z_{em}=1.18$
\citep{siemiginowska02,siemiginowska07}. In either case, this field
does not contain the typical isolated galaxy and absorber seen in many
quasar absorber fields.  \citep[e.g.,][]{sdp94,s97,gb97}.

The Q1127$-$145 field appears to have an unusually large number of
bright galaxies within $100''\times100''$ centered on the quasar line
of sight. Here we have performed a shallow spectroscopic survey of the
field at a limiting magnitude of $m_{F814W}\leq 20.3$ in an attempt to
identify the remaining absorbing galaxies within the field. At the
redshift of the DLA the $B-$band luminosity limit is
$L_B=0.15L_B^{\star}$.

\begin{table}
\begin{center}
  \caption{Summary of the galaxy spectroscopic observations obtained
    with the ARC 3.5m using the double imaging spectrograph
    (DIS). Four different DIS long-slit positions were obtained and
    their spatial are shown in Figure~\ref{fig:q1127field}. The
    instrument gratings and central wavelengths ($\lambda_c$) were selected to
    target H$\alpha$ and [\NII] emission lines, on the red channel,
    and {\OII} emission lines, on the blue channel, for $z\sim0.3$
    galaxies.}
  \vspace{-0.5em}
\label{tab:DIS}
{\footnotesize\begin{tabular}{lcccr}\hline
Slit &  & $\lambda_c$  &   & Exposure\\
Position &Grating & (\AA) &Date (UT) &(seconds)\\\hline
Slit 1 & B1200 &  4500          &2008 Jan. 16 &  10,500 \\
       & R1200 & 7300           &2008 Jan. 16 &  4500 \\
       & R1200 & 8500           &2008 Jan. 16 &  6000 \\
Slit 2 & B1200 & 5220           &2008 Feb. 01 &  4500 \\
       &  R830 & 9040           &2008 Feb. 01 &  4500 \\
Slit 3 & B1200 & 4700           &2008 Feb. 01 &  3000 \\
       & B1200 & 5220           &2008 Mar. 27 &  6000 \\
       &  R830 & 9040           &2008 Feb. 01 &  3000 \\
       &  R830 & 9040           &2008 Mar. 27 &  6000 \\
Slit 4 & B1200 & 5220           &2008 Feb. 01 &  3000  \\ 
       & R830 &  9040           &2008 Feb. 01 &  3000   \\\hline
\end{tabular}}
\end{center}
\end{table}

\subsection{Galaxy Spectroscopy}

Galaxy spectra were obtained during three nights between 2008 January
and 2008 March using the double imaging spectrograph (DIS) at the
Apache Point Observatory (APO) 3.5m telescope in New Mexico. Details
of the observations are presented in Table~\ref{tab:DIS}.  The
spectrograph has separate red and blue channels that have plate scales
of 0.40$''$~pixel$^{-1}$ and 0.42$''$~pixel$^{-1}$, respectively.  We
used a 1.5$''$-wide by 6$'$-long slit with no on-chip binning of the
CCD.

The B1200 grating was used for the blue channel resulting in a
spectral resolution of 0.62~\AA~pixel$^{-1}$ with wavelength coverage
of 1240~\AA. For the red channel, both the R830 and the R1200 gratings
were used. The R830 grating has a spectral resolution of
0.84~\AA~pixel$^{-1}$ with wavelength coverage of 1680~\AA. The R1200
has a spectral resolution of 0.58~\AA~pixel$^{-1}$ with wavelength
coverage of 1160~\AA.  Wavelength centers for each grating (see
Table~\ref{tab:DIS}) were selected to target {\OII}, $\rm{H} \alpha$,
and [\NII] emission lines for $z \sim 0.3$ galaxies.  The total
exposure time per target ranges from 3000 to 10,500 seconds and the
observations were performed during poor/cloudy weather conditions with
typical seeing of 1$-$2$''$. Four slit positions were obtained and are
shown in Figure~\ref{fig:q1127field}.
	
Spectra were reduced using IRAF\footnote{IRAF is written and supported
by the IRAF programming group at the National Optical Astronomy
Observatories (NOAO) in Tucson, Arizona. NOAO is operated by the
Association of Universities for Research in Astronomy (AURA), Inc.\
under cooperative agreement with the National Science Foundation.}.
External quartz dome-illuminated flat fields were used to eliminate
pixel-to-pixel sensitivity variations. Stellar spectra taken in the
same field were used as traces to facilitate the extraction of the
faint galaxy spectra.  Each spectrum was wavelength calibrated using
HeNeAr arc line lamps. The galaxy spectra were both vacuum and
heliocentric velocity corrected.


A Gaussian fitting algorithm \citep[see][]{archiveI}, which computes
best fit Gaussian amplitudes, centers, and widths, was used to obtain
the galaxy redshifts from one or more emission lines.  Emission lines
used to calculate the galaxy redshift were detected at or above the $3~\sigma$
level (the galaxy redshifts are listed in Table~\ref{tab:sample}).

Higher resolution spectra of three previously identified galaxies (G2,
G4, and G5) were obtained by \citet{kacprzak10}. Their Keck/ESI
spectra have a velocity resolution of $11$~\kms~pixel$^{-1}$ and have
a range of exposure times of 600--4200 seconds. Details regarding the
individual spectra and the data reductions are presented in
\citet{kacprzak10}. Here we present the flux calibrated spectra for
these three galaxies. The spectra were calibrated using IRAF with
standard stars taken during the night of the observation. We have made
no corrections for slit loss nor reddening.

\begin{table}
\begin{center}
  \caption{Summary of the imaging observations of the Q1127--145 field
    obtained using {\it HST\/} with the WFPC--2. The images were taken
    with the F814W filter with the quasar centered on chip 3 of the
    CCD array. The images were taken at a range of position angles
    (PA) listed below. In column 6 we list the Proposal IDentification
    number (PID) of the WFPC--2 observations taken by the PI
    Bechtold.}
\vspace{-0.5em}
\label{tab:HST}
{\footnotesize\begin{tabular}{ccrcccc}\hline
        & Quasar   & Chip & Exposure &     & \\
 Filter & Location & PA   & (seconds) & Date & PID       \\\hline
F814W  & WF3 &$-$11.502&4400 & May  23 2001&  9173 \\
F814W  & WF3 &$-$3.102 &4400 & Aug. 01 2001&  9173 \\
F814W  & WF3 & 147.006 &4400 & Nov. 16 2001&  9173 \\
F814W  & WF3 & 147.863 &4400 & Nov. 16 2001&  9173 \\
F814W  & WF3 & 164.698 &4400 & Jan. 03 2002&  9173    \\\hline
\end{tabular}}
\end{center}
\end{table}

\begin{table*}
\begin{center}
  \caption{Galaxy-absorber sample towards Q1127$-$145.  The table
    columns are (1) the galaxy ID as referenced in the text and
    figures, (2) the galaxy angular separation from the quasar line of
    sight, $\theta$, (3) the galaxy redshift, (4) the reference(s) for
    the galaxy spectroscopic identification, (5) the quasar-galaxy
    impact parameter, $D$, (6) the galaxy apparent magnitude, (7) the
    absolute B-band magnitude with its B-band luminosity (8), (9) the
    galaxy group ID, (10) {\MgII} absorption redshift, (11) the
    rest-frame {\MgII} $\lambda 2796$ equivalent width, $W_r(2796)$,
    and (12) the galaxy velocity offsets from the optical depth
    weighted mean {\MgII} absorption. Note that we find a group of at
    least five galaxies at similar redshifts as the DLA system
    ($z_{abs}=0.312710$) and two galaxies at similar redshifts to the
    weak {\MgII} absorption system at $z_{abs}=0.328266$. We also find
    five non-absorbing galaxies.}
\vspace{-0.5em}
\label{tab:sample}
{\footnotesize\begin{tabular}{lrccrcccccrr}\hline
 ID & $\theta$$\phantom{0}$        & $z_{gal}$ &REF$^a$ &  $D$$\phantom{0000}$  & $m_{F814W}$ & $M_B$ & $L_B$               &Group & $z_{abs}$ &W$_r(2796)$$^b$ &$\Delta v_{r}$$^c$$\phantom{00}$   \\
            &          $('')$ &           &                & (kpc)$\phantom{00}$ &             &       &       $(L_B^{\star})$ &  ID  &           & (\AA)$\phantom{0000}$  &          (\kms)  \\\hline
G1 & 3.81  & $0.3121$$\pm$$0.0003$    &  1      & $17.4$$\pm$$0.1$  & $21.55$$\pm$$0.37$ & $-$17.7  & 0.04 & 1  & 0.312710         & $1.773$$\pm$$0.006$& $+140$ \\
G2 & 10.01 & $0.3132$$\pm$$0.0002$    &  2,3,4  & $ 45.6$$\pm$$0.3$ &$ 18.81$$\pm$$0.11$ & $-$20.4 & 0.54 & 1  & 0.312710         & $1.773$$\pm$$0.006$& $-112$ \\
G3 & 16.23 & $0.32839$$\pm$$0.00003$  &  5      & $76.9$$\pm$$0.4$  &$ 20.12$$\pm$$0.20$ & $-$19.2 & 0.18 & 2  & 0.328266         & $0.029$$\pm$$0.003$& $-28$\\
G4 & 17.77 & $0.3124$$\pm$$0.0001  $  &  2,3,4  & $ 81.0$$\pm$$0.3$ &$ 18.64$$\pm$$0.10$ & $-$20.6 & 0.63 & 1  & 0.312710         & $1.773$$\pm$$0.006$& $+71$ \\
G5 & 19.30 & $0.32847$$\pm$$0.00003$  &  4      &$ 91.4$$\pm$$0.2$  &$ 18.84$$\pm$$0.11$ & $-$20.5 & 0.60 & 2  & 0.328266         & $0.029$$\pm$$0.003$& $-46$\\
G6 & 21.76 & $0.31167$$\pm$$0.00003$  &  5      & $99.8$$\pm$$0.1$  &$ 19.79$$\pm$$0.17$ & $-$19.4 & 0.22 & 1  & 0.312710         & $1.773$$\pm$$0.006$& $+238$ \\
G7 & 27.92 & $0.27921$$\pm$$0.00007$  &  5      & $118.3$$\pm$$0.8$ &$ 20.22$$\pm$$0.21$ & $-$18.7 & 0.11 & $\cdots$ &$\cdots$    & $<$0.0049      &$\cdots$  \\
G8 & 33.22 &$\cdots$$^d$          & $ \cdots$&$ \cdots $ & $20.08$$\pm$$0.19$ &$ \cdots$ & $\cdots$ &$\cdots$ & $\cdots$&  $\cdots$      &$\cdots$  \\
G9 & 33.91 & $0.20735$$\pm$$0.00006$  &   5     &$ 115.2$$\pm$$0.2$ &$ 19.85$$\pm$$0.19$ & $-$18.3 & 0.08 & $\cdots$ &$\cdots$    & $<$0.0050      & $\cdots$\\
G10& 37.93 &$\cdots$$^d$          & $ \cdots$&$\cdots$ & $19.94$$\pm$$0.19$ & $\cdots$ & $\cdots$&$\cdots$ &$\cdots$  &$\cdots$        &$\cdots$   \\
G11& 38.12 & $0.33293$$\pm$$0.00002$&    5     &$ 182.3$$\pm$$0.2$& $19.76$$\pm$$0.17$ & $-$19.6 & 0.27 &$\cdots$ &$\cdots$     & $<$0.0048      & $\cdots$ \\
G12& 43.23 & $0.30515$$\pm$$0.0004$$\phantom{0}$   &    5    &$ 195.0$$\pm$$0.6$ & $19.50$$\pm$$0.15$ & $-$19.65 & 0.27 &$\cdots$ &$\cdots$    & $<$0.0048      & $\cdots$  \\
G13& 50.08 & $\cdots$           & $ \cdots$ &   $\cdots$    & $20.14$$\pm$$0.20$ & $\cdots$ & $\cdots$&$\cdots$ & $\cdots$ & $\cdots$       &$\cdots$  \\
G14& 52.54 & $0.31243$$\pm$$0.00003$ &    5    &$ 240.8$$\pm$$0.3$  & $20.01$$\pm$$0.19$ & $-$19.2 & 0.18 & 1  & 0.312710         & $1.773$$\pm$$0.006$&$+64$  \\
G15& 68.22 & $0.2473$$\pm$$0.0002$   &    5    & $264.8$$\pm$$0.3$  & $19.00$$\pm$$0.12$ & $-$19.6 & 0.25 &$\cdots$ &$\cdots$     & $<$0.0057      &$\cdots$    \\\hline
\end{tabular}}
\end{center}
$^a$Galaxy Identification: (1)~\citet{lane98}, (2)~\citet{bb91},
(3)~\citet{gb97}, (4)~\citet{kacprzak10}, (5) This work.
$^b$Equivalent width limits are $3\sigma$.  $^c$$\Delta v_{r}$ is the
rest-frame velocity offset between the mean {\MgII} $\lambda 2976$
absorption line and the galaxy where, $\Delta
v_{r}=c(z_{abs}-z_{gal})/(1+z_{gal})$~\kms.  $^d$No strong emission
lines were detected.
\end{table*}

\subsection{Quasar Spectroscopy}

The absorption properties were measured from VLT/UVES \citep{dekker00}
archival spectra of Q1127$-$145 obtained on 2002 August 17 [PI Lane,
PID 67.A-0567(A)], 2003 August 18 [PI Savaglio, PID 69.A-0371(A)], and
2007 May 3 [PI Miniati, PID 076.A-0860(A)]. The UVES spectrum has a
wavelength coverage from 3046--4517~{\AA} and from
4622--6810~{\AA}. All spectra were taken with $2\times2$ binning using
a $1''$-wide slit, providing a spectral resolution with
FWHM~7\,km\,s$^{-1}$.  They were reduced using the standard ESO
pipeline and the custom code UVES Post--Pipeline Echelle Reduction
({\sc uves
popler}\footnote{http://astronomy.swin.edu.au/$\sim$mmurphy/UVES\_popler.html}). The
spectrum is both vacuum and heliocentric velocity corrected.  Analysis
of the {\MgII} absorption profiles was performed using interactive
software \citep[see][]{weakI,archiveI,cv01} for local continuum
fitting, objective feature identification, and measuring absorption
properties. The absorption redshifts are computed from the optical
depth weighted mean of the {\MgII} absorption profile
\citep[see][]{cv01}. The typical absorption redshift uncertainty is
$\sim0.3$~{\kms}.  Velocity widths of absorption systems are measured
between the pixels where the equivalent width per resolution element
recovers to the $1~\sigma$ detection threshold \citep{weakI}.

\subsection{HST Imaging}

The WFPC--2/{\it HST\/} F814W images were obtained from the Hubble
Legacy Archive (HLA\footnote{http://hla.stsci.edu/}) (PI
Bechtold). Details of the WFPC--2/{\it HST\/} observations are
presented in Table~\ref{tab:HST}.  Five sets of four 1100s exposures
were taken over a range of position angles. In all of the images, the
quasar was positioned in the center of chip 3 of the WFPC--2
camera.  We combined the five 4400s images using IRAF IMCOMBINE.

 Galaxy photometry was performed using the Source Extractor
(SExtractor) package \citep{bertin96} with a detection criterion of
1.5~$\sigma$ above background.  The $m_{F814W}$ magnitudes were
computed using the WFPC--2 zero points taken from Table 5.1 of the
WFPC--2 Data Handbook and the chip gains obtained from the WFPC--2
Instrument Handbook. All magnitudes are based upon the Vega system.

Galaxy absolute $B$-band magnitudes, $M_B$, were determined from the
$k$--corrected observed $m_{F814W}$.  The $k$--corrections were
computed using the formalism of \citet{kim96} using the spectral
energy distribution (SED) templates of \citet{kinney96}. We adopted a
Sb SED which is consistent with the average colour of {\MgII}
absorbing galaxies \citep{sdp94,zibetti07}.  $B$--band luminosities
were computed using the DEEP2 optimal $M^{\star}_B$ of
\citet[][Table~2]{faber07} in the redshift bin appropriate for each
galaxy ($M^{\star}_B=-21.07$ for $\left< z \right> =0.3$).

\section{Results}
\label{sec:results}

\begin{figure*}
\includegraphics[angle=0,scale=1.0]{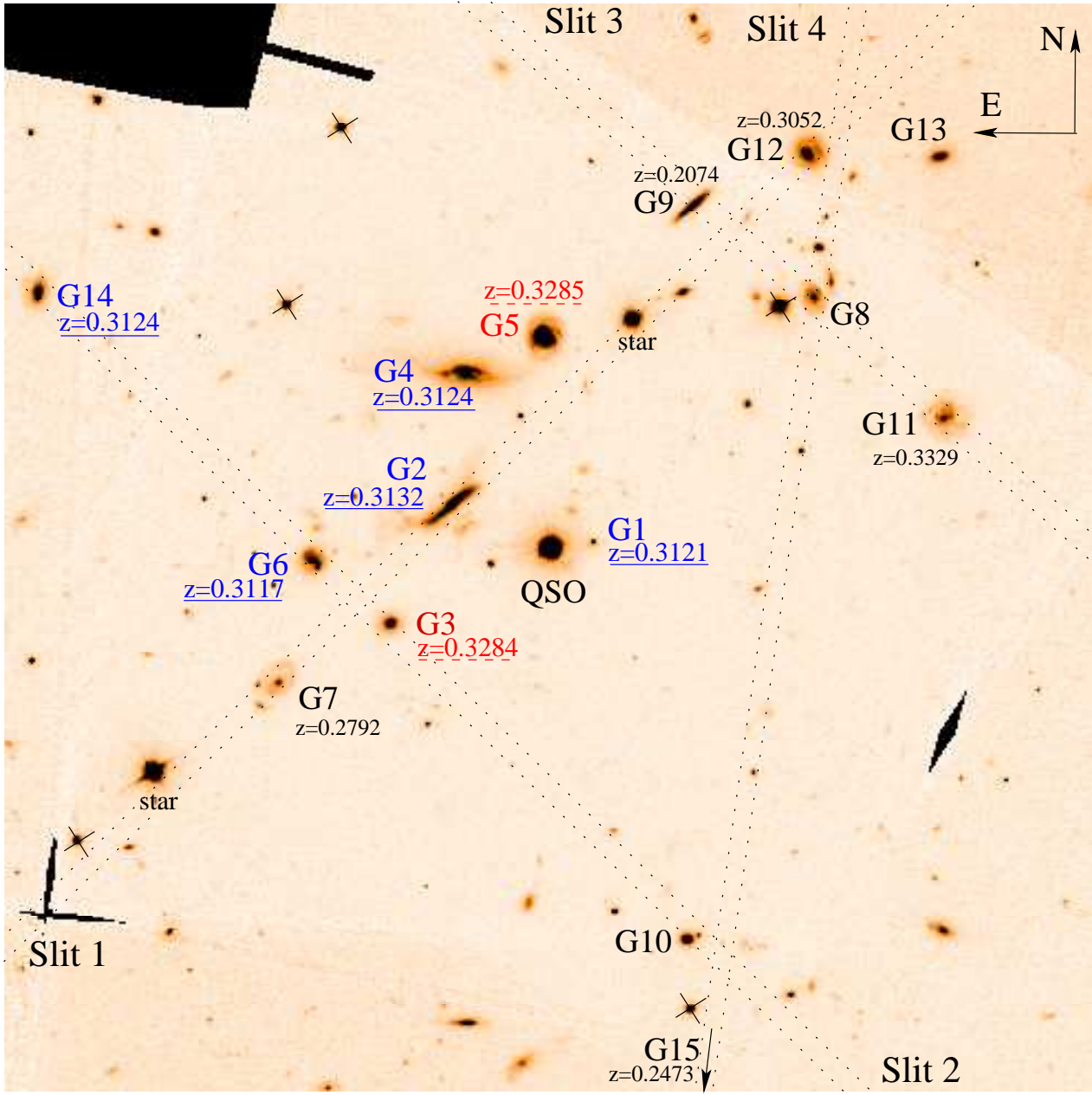}
\caption{$100'' \times 100''$ {\it HST}/WFPC--2 F814W image of the
quasar field Q1127$-$145. The quasar is located in the center of the
image. Black regions have no WFPC--2 coverage. The four slit positions
are indicated with dotted lines. All galaxies with spectroscopically
confirmed redshifts are indicated. The galaxy redshifts are only
quoted to four significant figures for clarity (see
Table~\ref{tab:sample} for full redshifts). Galaxy G15, an image of
which can be found in Figure~\ref{fig:abs0312}, is beyond the $100''
\times 100''$ image shown here.  Note there are two distinct groupings
of galaxies -- one at $z\sim0.313$ (blue/solid line) and at
$z\sim0.328$ (red/dashed line) -- and the former is the galaxy group
associated with the DLA at $z_{abs}=0.313$. Point source objects,
which are likely stars, are indicated with an ``X''. Objects that have
been spectroscopically identified as stars are indicated.}
\label{fig:q1127field}
\end{figure*}

\begin{figure*}
\includegraphics[angle=0,scale=0.30]{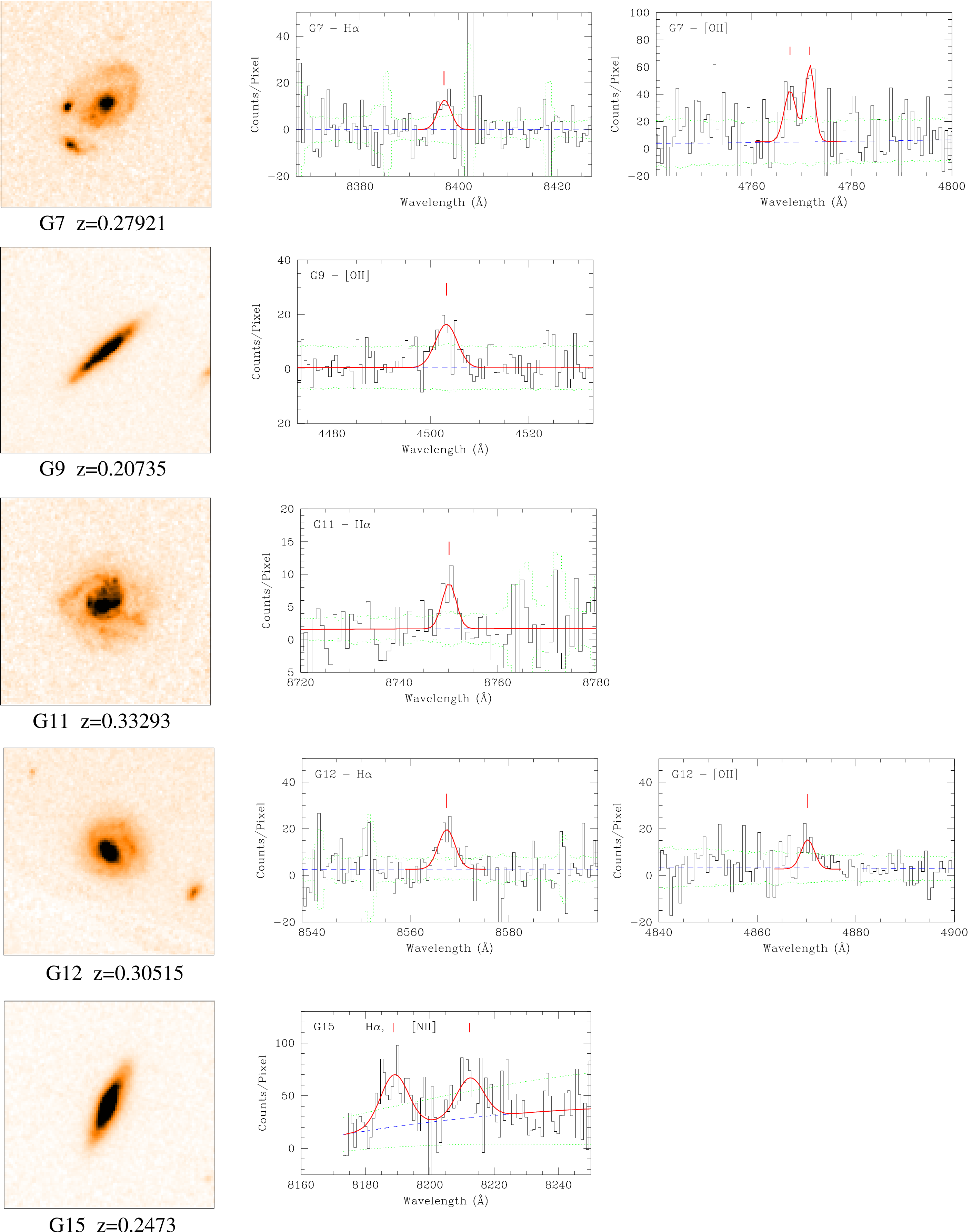}
\caption{ $10\times10''$ WFPC--2 F814W images of the non-absorbing
  galaxies G7 (image size $42.4\times42.4$~kpc), G9
  ($34.0\times34.0$~kpc), G11 ($47.8\times47.8$~kpc), G12
  ($45.1\times45.1$~kpc), and G15 ($38.8\times38.8$~kpc). The galaxies
  are shown in increasing impact parameter order and have the same
  orientation as in Figure~\ref{fig:q1127field}. The H$\alpha$ and/or
  {\OII} galaxy emission lines used to spectroscopically confirm the
  galaxy redshift are shown. Three galaxies (G7, G12, and G15) have
  been identified using two emission lines while two galaxies (G9 and
  G11) have been identified with one. The blue dashed lines are the
  fits to the continuum and the solid red curves are fits to the
  data. The green dotted lines are the $3\sigma$ detection limits. }
\label{fig:nonabs}
\end{figure*}

Here we discuss the galaxies identified in our redshift survey along
with galaxies identified in previous works. In
Figure~\ref{fig:q1127field} we present a 100$\times$100$''$ portion of
the combined WFPC2 image centered on the quasar. The four slit
positions used in our new observations are indicated on the image.  In
Table~\ref{tab:sample}, we list all the galaxies in the quasar field
that have $m_{F814W}<20.3$ within $100\times100''$ box centered on the
quasar (we have included galaxy G1 since it has a spectroscopic
redshift and G15 which is beyond the surveyed region).

We have obtained spectroscopic redshifts for eight new galaxies in
this work.  We have identified: (1) a group of galaxies associated
with the DLA, (2) a pair of galaxies associated with a weak {\MgII}
absorption system, and (3) five non-absorbing galaxies.  The offsets
of the systemic velocity of the absorbing galaxies from the optical
depth weighted mean {\MgII} absorption are also listed in
Table~\ref{tab:sample} and range from $-112$ to $+238$~{\kms}.  In the
following subsections we will discuss the galaxies identified in the
Q1127$-$145 field.

\subsection{Non-Absorbing Galaxies}

We have spectroscopically confirmed the redshifts of five
non-absorbing galaxies for which we do not detect {\MgII} absorption
to the limits of the UVES spectrum. In Figure~\ref{fig:nonabs} we show
$10''\times10''$ images of these galaxies along with their emission
line spectra. Galaxies are listed in increasing impact parameter
order. The galaxy redshifts were determined using H$\alpha$ and/or
{\OII} emission lines.  From the {\it HST} image, the non-absorbing
galaxies appear to be normal spiral disks and the spectra indicate
they have ongoing star formation.

Three galaxies (G7, G12, and G15) have been spectroscopically
confirmed with two emission lines while two galaxies (G9 and G11) have
been confirmed with a single line.  There is the possibility that the
redshifts of galaxies computed with only one emission line may be
incorrect. However, given the wavelengths of these lines and the
observed apparent magnitudes of the galaxies, it is highly unlikely
that these are at different redshifts than the ones quoted here. The
redshift of G9 was computed using the {\OII} emission line and is
reliable because since the observed {\OII} wavelength is
$\sim4503$~{\AA}, which is bluer than any other optical galaxy
emission line rest wavelength and is not likely to be a UV emission
line since it would place the galaxy at a redshift $z>2$. The
redshifts of G11 and G14 were computed using only the H$\alpha$
emission line. If the H$\alpha$ emission line is incorrectly
identified then the next likely candidate line would be {\OIII}. This
would place the galaxy at redshifts of $z>0.7$, which would result in
$L_B \geq 1.5 L_B^\ast$ with a disk scale length $\geq 4.5$~kpc. We
are therefore confident in these emission line identifications.

All non-absorbing galaxies have {\MgII} equivalent width 3$\sigma$
detection limits of 4.8$-$5.7~m{\AA} (see
Table~\ref{tab:sample}). These equivalent width limits are quite low
and imply that these galaxies are not associated with any substantial
{\MgII} absorption along the LOS, which can be interpreted as either
the quasar LOS passing outside the galaxies' {\MgII} enriched halos or
through a void within the patchy {\MgII} halo gas distribution. It is
important to note that all the non-absorbing galaxies have impact
parameters $D>118$~kpc.  This is consistent with current studies that
show most luminous galaxies at projected distances within
$D\sim$120~kpc are {\MgII} absorbers while beyond $D\geq$120~kpc they
are not \citep{cwc-china,zibetti07,kacprzak08,chen08}.

\begin{figure}
\includegraphics[angle=0,scale=0.30]{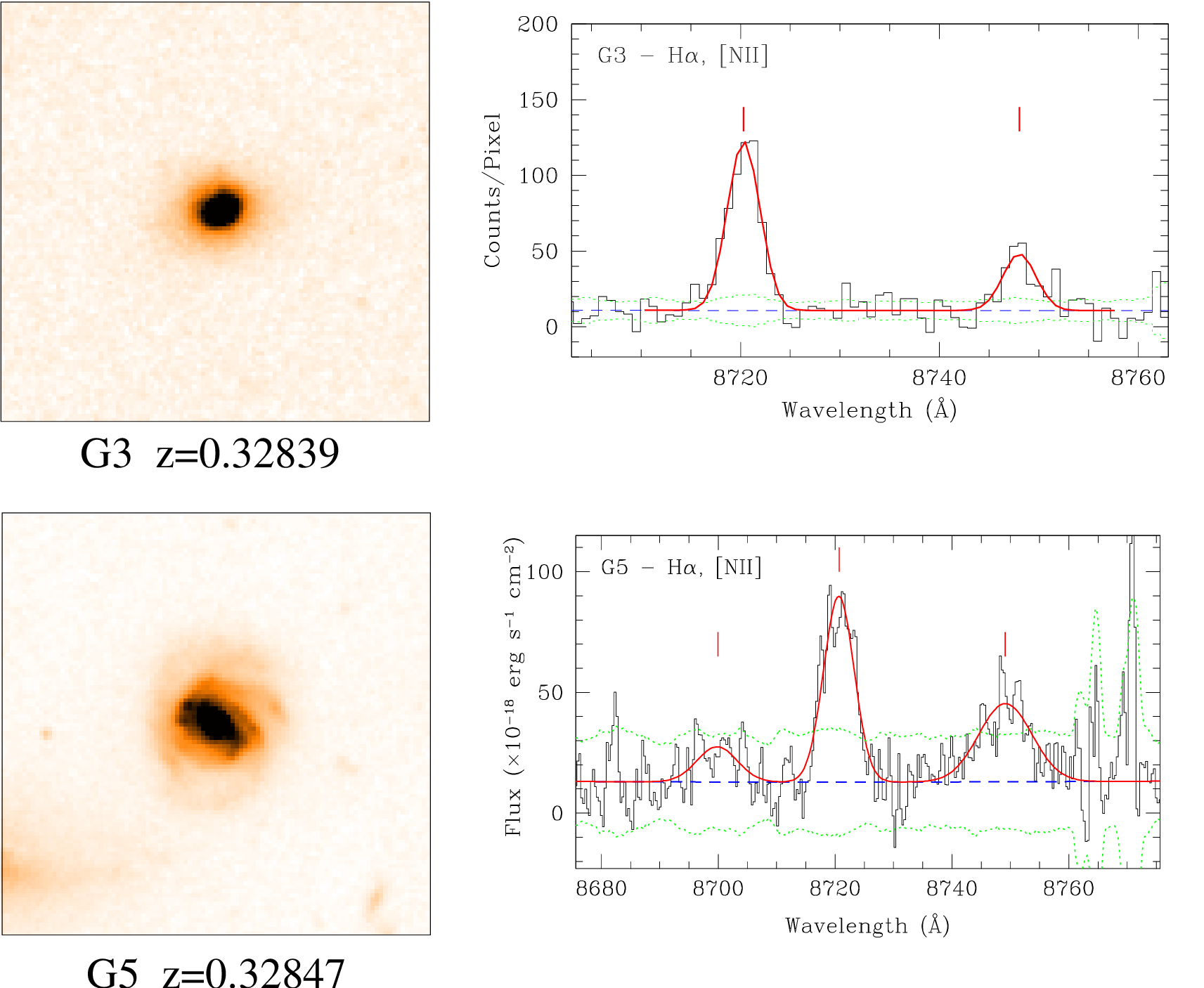}
\caption{ Same as Figure~\ref{fig:nonabs} except for the two galaxies
G3 and G5 associated with the $z=0.328$ absorption system. The
spectrum of G5 was obtained with Keck/ESI and is flux calibrated. The
images are 47.4$\times $47.4~kpc in size at the redshift of the
absorption.}
\label{fig:abs0328}
\end{figure}

\subsection{$z_{abs}=0.328$ Galaxies}

We have spectroscopically confirmed one new galaxy (G3) at $z=0.328$
in addition to G5, which was spectroscopically identified by
\citet{kacprzak10}. Both galaxies have strong H$\alpha$ and {\NII}
emission lines shown in Figure~\ref{fig:abs0328}. G3 has the smallest
impact parameter of $D=76.9$~kpc and has a H$\alpha$ rest equivalent
width of 32.0~{\AA}. G5 has an impact parameter of $D=91.4$~kpc and
has a H$\alpha$ rest equivalent width of 29.5~{\AA}. Both galaxies
appear to have normal unperturbed morphology. G3 is a compact core
0.2$L_B^{\star}$ galaxy with ongoing star formation while G9 is a
0.6$L_B^{\star}$ galaxy with a large bar and a sizable bright bulge
similar to a local SBb galaxy. The major differences between the two
galaxies are that G9 is much brighter than G3 and exhibits spiral
arms.

\begin{figure}
\includegraphics[angle=0,scale=0.92]{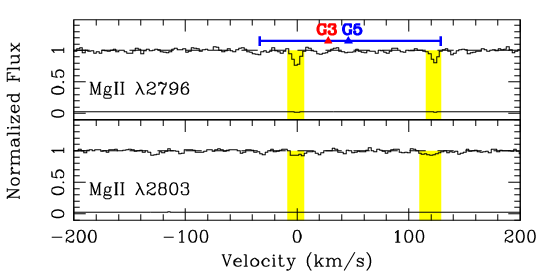}
\caption{ The observed {\MgIIdblt} absorption in the UVES quasar
  spectrum at velocities relative to $z_{abs}=$0.328266.  The dashed
  line is a fit to the continuum and the solid line near zero is the
  flux error spectrum.  The \MgII $\lambda2796$ absorption redshift
  is the zeropoint of the velocity scale and the absorption velocities
  are shaded in. The \MgII $\lambda2796$ is detected to a
  significance level of 5~$\sigma$ and the \MgII $\lambda2803$ is
  detected to a significance level of 3~$\sigma$. The redshifts of
  both galaxies are indicated by the triangles and the maximum
  observed rotation velocities of G5, obtained from
  \citet{kacprzak10}, are shown by the error bars. Note that if one
  includes the kinematics of both galaxies, they both have velocities
  consistent with the {\MgII} absorption.}
\label{fig:gasabs0328}
\end{figure}

The $z_{abs}=0.328$ {\MgII} absorption was first reported by
\citet{narayanan07}, and is presented in
Figure~\ref{fig:gasabs0328}. This weak system is composed of two
single clouds with a velocity separation of 124~{\kms} and has a total
equivalent width of $W_r(2796)=0.029$~{\AA}. No significant {\MgI} is
detected ($3\sigma$, $W_r\leq0.003$~\AA). In
Figure~\ref{fig:gasabs0328} we also show the galaxy systemic redshifts
(triangles) relative to the absorption system. Note that both galaxies
have redshifts that are bracketed with both absorption clouds.

Galaxy G5's rotation curve was obtained by \citet{kacprzak10} and the
error bars in Figure~\ref{fig:gasabs0328} indicate its observed
maximum rotation velocities. With a projected velocity span of
$\sim$160~{\kms}, G5's dynamics is consistent with both absorption
cloud velocities. In particular, \citet{kacprzak10} show that the two
cloud velocities align with each side of the rotation curve, but that
pure disk models are unable to reproduce the observed {\MgII}
absorption velocities. If we were able to obtain a rotation curve for
G3, both galaxy kinematics would likely be consistent with both
absorption clouds. Thus, in these circumstances, it is difficult to
disentangle which galaxy is associated with this particular
absorber. Yet associating the absorption with one galaxy or the other
may result in a difference in conclusions regarding the origins of the
absorption. We discuss the implications of this in Section~4.

\subsection{$z_{abs}=0.313$ Galaxies}

\begin{figure*}
\includegraphics[angle=0,scale=0.30]{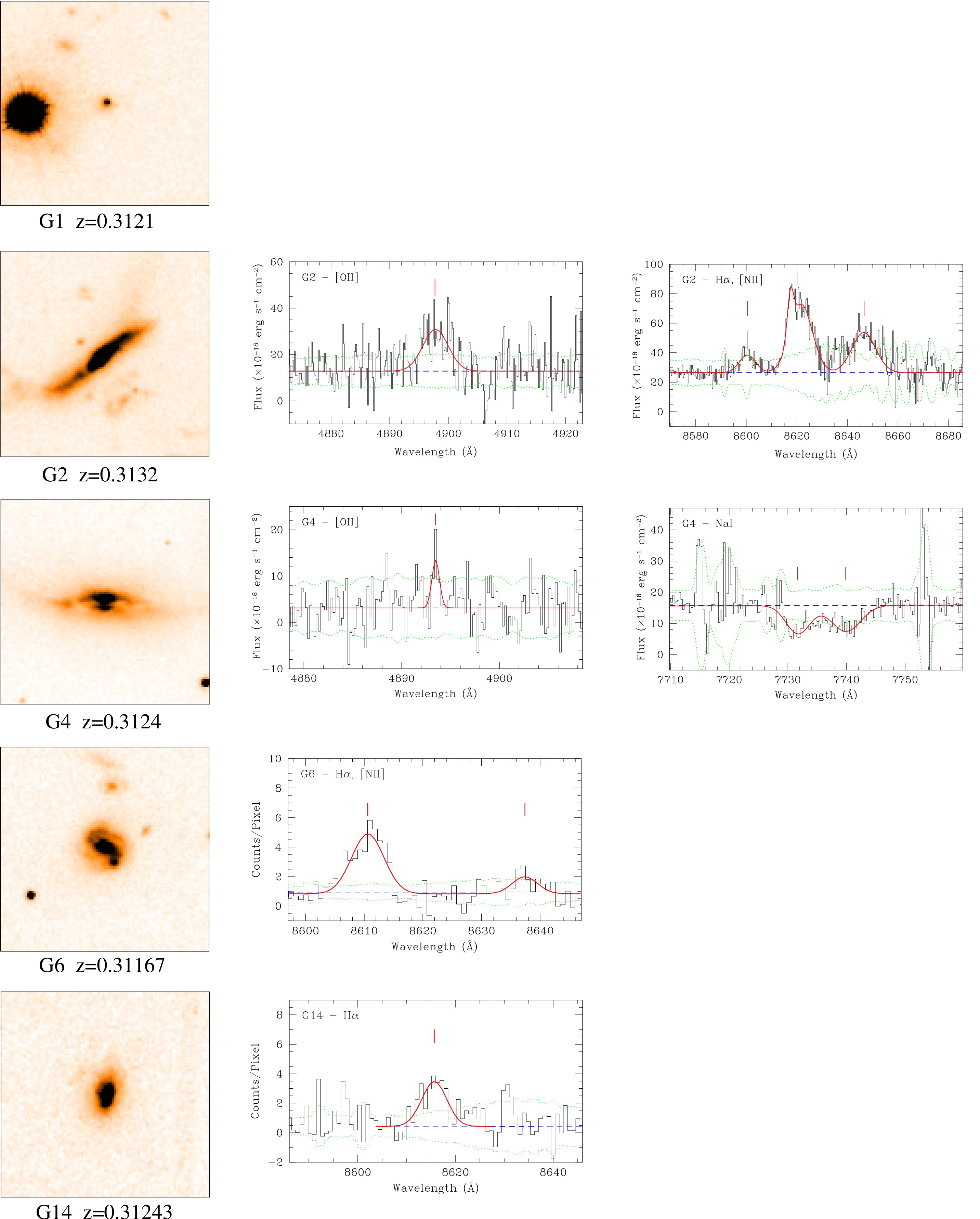}
\caption{ Same as Figure~\ref{fig:nonabs} except for the G1, G2, G4,
G6, and G14 galaxy group associated with the $z=0.313$ DLA.
 G1 was spectroscopically confirmed by
\citet{lane98} and we do not have a spectrum of this galaxy to show
here. The spectra of G2 and G4 were obtained with Keck/ESI and are
flux calibrated. The images are 45.9$\times $45.9~kpc in size at the
redshift of the DLA absorption. For G4, we show the {\NaI} absorption
feature for reference since it was also used to confirm the redshift
since the {\OII} line is quite weak.}
\label{fig:abs0312}
\end{figure*}

\begin{figure*}
\includegraphics[angle=0,scale=0.92]{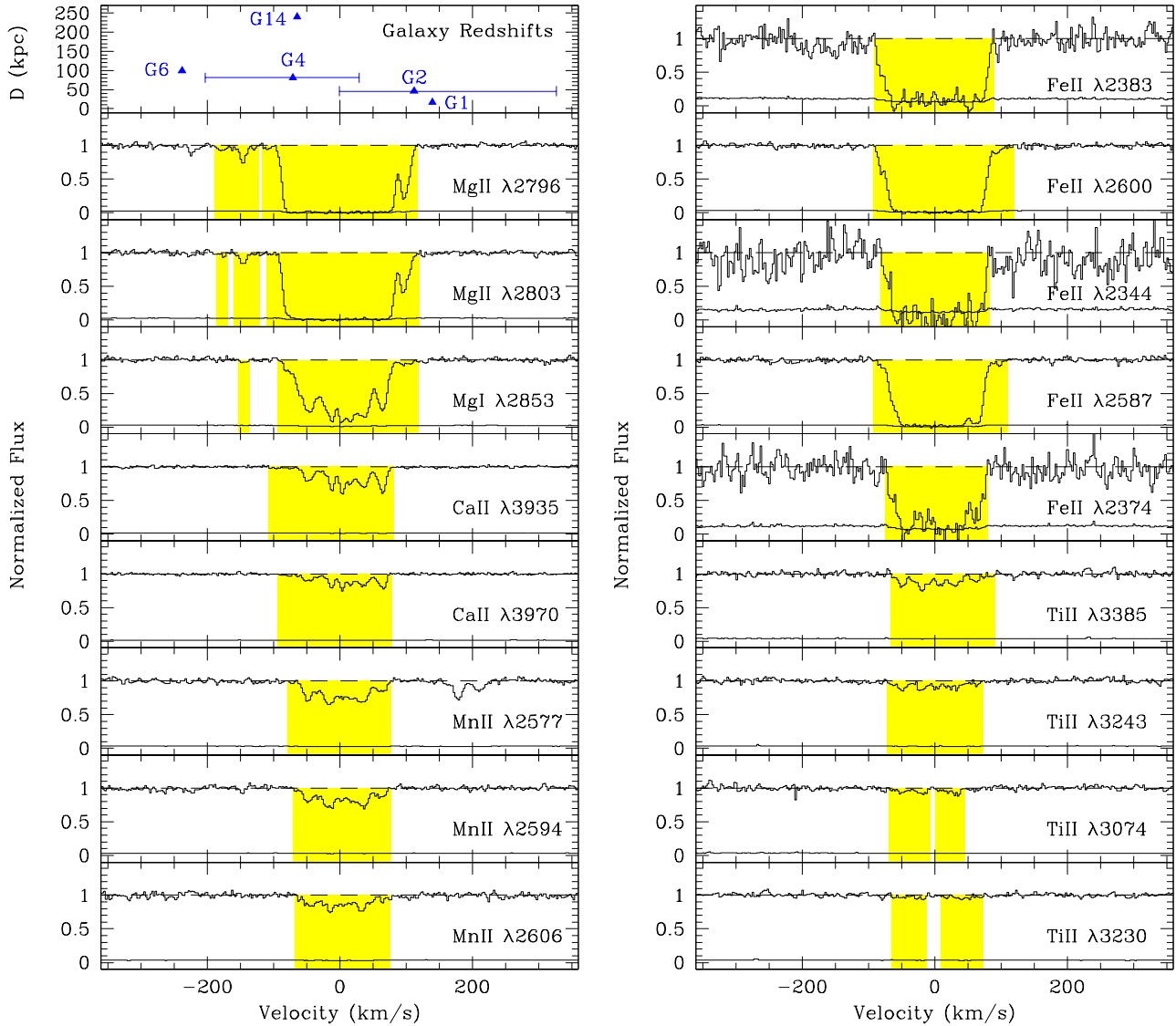}
\caption{ The observed metal-line absorption from the $z_{abs}=0.313$
  DLA obtained from the UVES quasar spectrum.  The dashed line is a
  fit to the continuum and the solid line near zero is the flux error
  spectrum. The \MgII $\lambda2796$ absorption redshift is the
  zeropoint of the velocity scale. The range of velocities over which
  significant absorption is detected are shaded for each transition;
  for \MgII $\lambda2796$, the detection threshold was set at
  5$\sigma$ while a lower threshold of 3$\sigma$ is set for all other
  transitions.  The redshifts of all five galaxy group members are
  indicated by the triangles and are shown as a function of impact
  parameter. The maximum observed rotation velocities of G2 and G4
  obtained from \citet{kacprzak10} are shown by the error bars. Note
  the redshift distribution of the galaxy group spans all of the
  absorption velocities. The observed rotational kinematics of G2 and
  G4 alone also span the absorption velocities. }
\label{fig:gasabs0312}
\end{figure*}

\begin{figure*}
\includegraphics[angle=0,scale=0.82]{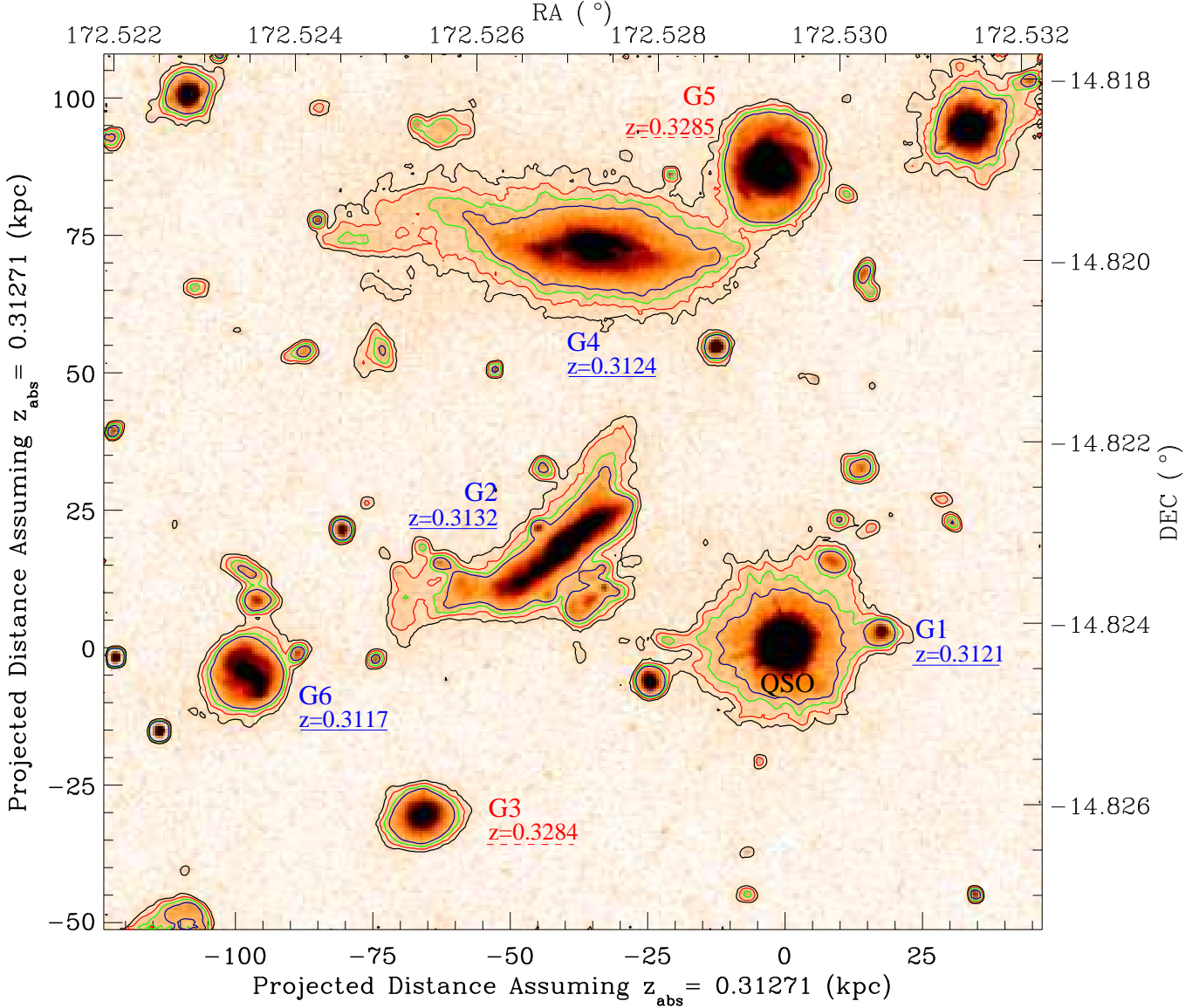}
\caption{Smoothed surface brightness contour plot of the quasar
  field. The contours are in 0.3 mag arcsec$^{-2}$ intervals:
  23.3(blue), 23.6(green), 23.9(red), and 24.2(black) mag
  arcsec$^{-2}$. The surface brightness limit of the image is roughly
  25 mag arcsec$^{-2}$. The axes indicate the projected distance away
  from the quasar in kpc (assuming $z=z_{abs}=0.31271$) 
   along with the RA and DEC in units of
  degrees.  Note the extend optical streams - possibly tidal tails -- seen for G2 and G4.}
\label{fig:q1127cont}
\end{figure*}

\begin{table*}
\begin{center}
  \caption{The emission line rest frame equivalent widths and emission
    line fluxes for the absorbing galaxies. The {\OII} equivalent
    width was not published by \citet{lane98}. Only the Keck/ESI
    spectra of G5, G2, and G4 are flux calibrated. }
\vspace{-0.5em}
\label{tab:flux}
{\footnotesize\begin{tabular}{llcccccccccccccc}\hline
& \multicolumn{9}{c}{Rest Equivalent Width ({\AA})} & \multicolumn{5}{c}{Integrated Line Flux ($\times 10^{-18}$~erg~s$^{-1}$~cm$^{-2}$)}\\\cline{3-8}\cline{10-15}
                 &                 &       &       &           &        &   &      &          &       &        &           &         \\[-1.5ex]
ID& $z_{gal}$& \OII & H$\beta$ & \OIII &[\NII] & H$\alpha$ & [\NII] &   & \OII & H$\beta$ & \OIII & [\NII] & H$\alpha$ & [\NII]  \\\hline

G3 & 0.32839 &$<$5.1&$\cdots$&$\cdots$ &$<$3.7&$32.0$$\pm$$1.9$ &$11.0$$\pm$$1.1$&  &$\cdots$& $\cdots$ &$\cdots$&$\cdots$ &$\cdots$&$\cdots$  \\
G5 & 0.32847 &$<$2.0&$<$2.0&$<$1.2 &$7.4$$\pm$$1.8$&$29.5$$\pm$$2.3$ &$22.1$$\pm$$2.3$&  &$<$76& $<$43 &$<$23&$128$$\pm$$31$ &$511$$\pm$$30$&$382$$\pm$$36$  \\\hline
\multicolumn{15}{c}{$z_{abs}=0.313$ Group}\\\hline
G1 & 0.3121  &$\cdots$ &$\cdots$&$\cdots$ &$\cdots$&$\cdots$ &$\cdots$& &$\cdots$& $\cdots$ &$\cdots$&$\cdots$ &$\cdots$&$\cdots$  \\
G2 & 0.3132  &$5.8$$\pm$$0.5$ &$<$0.4&$<$0.4 &$2.7$$\pm$$0.4$&$17.3$$\pm$$1.3$ &$\phantom{0}$$9.0$$\pm$$0.8$& &$ 99$$\pm$$ 8$& $<$11 &$<$12&$95$$\pm$$15$ &$602$$\pm$$33$&$295$$ \pm$$ 28$  \\  
G4 & 0.3124  &$2.4$$\pm$$1.0$ &$<$0.5&$<$0.6 &$<$1.4&$<$1.5 &$<$1.0& &$10$$\pm$$4$& $<$7  &$<$8 &$<$31 &$<$34& $<$23  \\
G6 & 0.31167 &$<$4.2   &$\cdots$&$\cdots$ &$<$2.9 &$32.4$$\pm$$1.9$ &$11.1$$\pm$$1.1$& &$\cdots$& $\cdots$ &$\cdots$&$\cdots$ &$\cdots$&$\cdots$  \\ 
G14& 0.31243 &$\cdots$ &$\cdots$&$\cdots$ &$<$2.5&$34.8$$\pm$$3.9$ &$<$5.1& &$\cdots$& $\cdots$ &$\cdots$&$\cdots$ &$\cdots$&$\cdots$   \\\hline
\end{tabular}}
\end{center}
\end{table*}

Three previously spectroscopically confirmed galaxies associated with
the DLA are G1, G2, and G4. G1 is the closest galaxy to the quasar
line of sight having an impact parameter of $D=17.4$~kpc. This faint
0.04$L_B^{\star}$ compact core galaxy was identified from {\OIII}
emission lines \citep{lane98}. The next closest galaxy to the quasar
line of sight is G2 with $D=45.6$~kpc. This $0.54L_B^\star$ edge-on
spiral displays asymmetries on both sides of the galaxy and its
H$\alpha$ rest equivalent width is 17.3~{\AA}.  The G4 galaxy has a
major dust lane and a large bulge. Given that we detected no strong
emission lines, this galaxy could either be classified as Sa or as an
early-type S0 galaxy. It has a luminosity of $L_B=0.63L_B^\star$ with
impact parameter of $D=81.0$~kpc.

We have spectroscopically confirmed two new galaxies (G6 and G14) at a
similar redshift to the $z_{abs}=0.313$ DLA in addition to the three
previously discovered galaxies G1, G2, and G4 \citep{bb91,lane98}. All
five galaxies are shown in increasing impact parameter order in
Figure~\ref{fig:abs0312}. The galaxy emission line strengths are shown
in Table~\ref{tab:flux}.  The newly spectroscopically confirmed galaxy
G6 has an impact parameter of $D=99.8$~kpc and has a H$\alpha$ rest
equivalent width of 32.4~{\AA}. This 0.22$L_B^{\star}$ spiral galaxy has
asymmetric spiral arms. G14 has an impact parameter of $D=240.8$~kpc
and has a H$\alpha$ rest equivalent width 34.8~{\AA}. This
0.18$L_B^{\star}$ galaxy has a perturbed early-type morphology. Both
newly identified galaxies are at larger impact parameters than the previously 
known ones (G1, G2 and G4), suggesting that the galaxy
environment is more group-like and much larger than previously
thought.

The galaxy group has five confirmed galaxies that have a
luminosity range of $0.04\leq L_B\leq0.63 L_B^{\star}$. The group of at
least five galaxies has a velocity dispersion of $\sigma=115$~{\kms}
centered at a redshift of $z_g=0.31236$. 

The velocity spread of the galaxy group is comparable to the velocity
spread of the {\MgII} gas as seen in Figure~\ref{fig:gasabs0312}.  The
galaxies cover a full velocity range of 350~{\kms}.  The {\MgII}
absorption seems to occur in two separate kinematic components: the
large, saturated component has a large velocity spread of 235~\kms,
and the much weaker absorption bluewards of the saturated component
comprises several clouds with a velocity spread of 68~\kms.  The
{\MgII} absorption redshift is offset $80$~{\kms} redward of the
galaxy group redshift of $z_g=0.31236$.  In
Figure~\ref{fig:gasabs0312}, we show the rotation velocity ranges for
G2 and G4 from \citet{kacprzak10}. These galaxies are the brightest
two in the group. The maximum observed projected rotational velocity
for G2 is $204$~{\kms} and G4 has maximum observed projected
rotational velocity of $90$~{\kms}.  The rotation velocities of both
G2 and G4 alone cover the full range of the absorption
velocities. \citet{kacprzak10} note that, using a simple disk model, a
large fraction of the absorption velocities could be explained by the
halo gas of G2 and G4 rotating as thick disks. However, the disk model
is quite unrealistic since it assumes the halo rotates at constant
velocity, set by the maximum galaxy rotation velocity, independent of
scale height. Therefore it is unlikely that all of the absorption is
produced by co-rotating halo gas from both galaxies G2 and G4.

Using the {\MgI} profile, one can study the individual clouds that are
saturated in {\MgII}. The majority of the {\MgI} gas is aligned with
the saturated {\MgII} component. Only a very weak {\MgI} cloud is
detected in a second kinematic component at $-$150~\kms. All
detectable {\FeII} lines are highly saturated (see
Figure~\ref{fig:gasabs0312}). The absorption velocity structure is
more apparent in the {\CaII}, {\MnII} and {\TiII} lines where five to
six kinematically distinct absorption features are apparent. These
features do not align with the five galaxy redshifts in the group,
making it difficult to determine the origin of the absorbing gas.

A {\it HST}/STIS E230M spectrum was also taken of this quasar
(PI:Bechtold). The {\ZnII} present in the spectrum has the same
velocity structure as the {\CaII} and the measured {\ZnII} column
density is $13.45\pm0.08$~\cmsq (Kanekar 2009, private
communication). We discuss the metallicity of this DLA in greater
detail in Section~\ref{sec:metals}.

\citet{lane98} detected 21-cm absorption at a similar redshift to the
DLA. Higher signal-to-noise and higher resolution 21-cm absorption
spectra taken by \citet{chengalur00} revealed that the broad feature
detected by \citet{lane98} breaks up into a five to six narrow
components. The full velocity width of the absorption is $\sim
120$~{\kms}, which is comparable to the velocity widths of the
{\CaII}, {\TiII} and {\ZnII}.  The individual 21-cm absorption
components also have a similar velocities as the heavy metal
absorption components. The 21-cm profile was found to vary on short
timescales of a few days and models that reproduce the variable 21-cm
absorption profile required small-scale variations of the optical
depth of the absorber \citep{kanekar01}.


\subsection{Galaxy Morphologies}

In Figure~\ref{fig:q1127cont} we show a zoomed-in contour plot of the
galaxies in the field. The axes are set to the physical scale at the
redshift of the DLA. The WFPC--2 F814W filter is comparable to the
rest frame Kron-Cousins R filter at the $z=0.313$ absorption redshift.
Four of the galaxies in the $z=0.313$ group are visible here along
with both $z=0.328$ galaxies. We plot surface brightness contours
ranging between $23.3-24.2$~mag~arsec$^{-2}$. The image has a surface
brightness limit of $\sim25$~mag~arcsec$^{-2}$.

Upon inspecting the galaxies in the $z=0.313$ group, one notices that
both G2 and G4 seem to exhibit tidal disturbances. G2 contains either
a strong warp in the disk or tidal tails from previous
interactions. The rest-frame $R$-band optical streams extend up to 25
kpc away from the galaxy. These features suggest at least one
merger/harassment event. The small galaxy situated below the
semi-major axis of G2 in projection does not have a spectroscopic
redshift. Using broad band photometry, \citet{chen03} derive a
redshift of $z=0.53$ for this faint object. However, if this object is
undergoing an interaction with G2, then its colours may not be
consistent with the standard spectral templates used to derive the
photometric redshift. The redshift of this galaxy is in need of
spectroscopic confirmation.

The S0-like galaxy G4, exhibits a strong dust lane and also has an
extended optical stream extending $\sim$30 kpc projected in length
towards the East from the galaxy center. This potential tidal stream
is suggestive of previous interactions, possibly with G2. There are
also several unidentified, large, low surface-brightness blobs above
and below the stream, which may be associated with the tidal debris or
high-redshift low surface brightness galaxies.  The asymmetric spiral
G6 may have some tidal debris as well since it has some optical
structures in its vicinity but they could potentially be galaxies at
different redshifts.  The morphologies of the galaxies seen here
suggest that this group has undergone some interactions in the past.

It has been noted in the literature \citep{rao03}, and can be seen
here, that there appears to be a significant level of surface
brightness (SB) around the quasar. \citet{rao03} suggest that the low
SB feature is potentially a low SB galaxy that is responsible for the
DLA absorption.  However, no evidence yet exists to support this
possibility. It also remains possible that the emission seen in
Figure~\ref{fig:q1127cont} is associated with the $z_{em}=1.18$ quasar
host galaxy, which has extended radio and x-ray emission that overlap
quite well with the extended quasar optical emission
\citep[see][]{siemiginowska02,siemiginowska07}.  There are many faint
galaxies/structures that remain unidentified in this field, several of
which reside in close proximity to the quasar line of sight, that may
also contribute to the absorption.

Our data suggest that most galaxies in the $z=0.313$ group have
undergone interactions in the past.  One way to differentiate between
these and the many possible origin scenarios for the DLA gas is by
studying the galaxy star formation rates and by comparing the
metallicity of the surrounding galaxies and of the absorption.

\subsection{Metallicities and Star Formation Rates}
\label{sec:metals}

In an effort to better understand the origins of the metal-enriched
absorption, we compute, when possible, the galaxy star formation rates
(SFRs) and metallicities.

We can only compute SFRs for galaxies that have Keck/ESI spectra which
have been flux calibrated. The APO/DIS data were taken during
poor/cloudy weather conditions and we are unable to flux calibrate
them. We compute the galaxy SFRs using the H$\alpha$ \citep{kewley02}
and {\OII} \citep{kewley04} emission line relations. We do not apply
any dust corrections since we are unable to measure the Balmer
decrement. We do not apply slit loss corrections.  Thus, the SFRs
quoted here are lower limits.

By contrast with the SFR calculations, we are able to compute
metallicities for additional galaxies using only the equivalent widths
of H$\alpha$ and {\NII} emission lines.  Since both H$\alpha$ and
{\NII} are only $20.66$~{\AA} (at rest wavelengths) apart, the
continuum flux levels are approximately the same and are insensitive
to dust reddening. Thus, the metallicity indicator
$N2=f($\NII$\lambda6583)/f($H$\alpha)$, which is a ratio of emission
line fluxes ($f$), becomes just the ratio of equivalent widths. This
technique has been demonstrated to work for other metallicity
indicators such as $R_{23}$ \citep[see][]{kobulnicky03}.  We apply the
$N2$ metallicity relation from \citet{pettini04}, where
12$+$log(O/H)$=8.90+0.57\times N2$. Note that the $N2$ metallicity
indicator becomes unreliable above roughly solar since the $N2$ index
saturates as nitrogen becomes the dominant coolant \cite[see][ and
references therein]{erb06}.  We assume a solar oxygen abundance of
log(O/H)$_{\odot} =8.736\pm0.078$ \citep{holweger01}. The star
formation rates and metallicities are listed in Table~\ref{tab:SFR}.

For the $z=0.313$ group we are able to measure star formation rates
for galaxies G2 and G4. For G2 we compute
SFR({\OII})=0.44~M$_{\odot}$~yr$^{-1}$ and
SFR(H$\alpha$)=1.52~M$_{\odot}$~yr$^{-1}$. For G4 we compute
SFR({\OII})=0.045~M$_{\odot}$~yr$^{-1}$. Again, we have not applied
any dust corrections which is probably the source of the difference
between the SFRs derived from H$\alpha$ and {\OII}; the H$\alpha$ SFR
is more reliable since it is less affected by dust extinction than
{\OII}.  Since the galaxy SFRs are not corrected for dust extinction,
the SFRs quoted are lower limits. We note that both of these galaxies
have typical SFRs and are not likely to have strong winds. Even if
dust corrections were applied, it would at most increase the SFRs by a
factor of $\sim 2$ which is still lower than expected for galaxies
with strong winds. The S0-like morphology of G4 is consistent with low
SFRs and no strong outflow winds. Given that G2 and G4 are the
brightest galaxy group members closest to the quasar LOS, it is
unlikely that the absorption is coming from winds.  We can compute the
metallicities for three of the galaxies in the group: [O/H]$=0.00\pm
0.09$ for G2, [O/H]$=-0.10\pm 0.09$ for G6, and a limit of [O/H]$<
-0.31$ for G14. These galaxies are roughly solar in abundance except
for G14 which is less than 1/2 solar.

The DLA absorption metallicity was initially derived from the amount
of photoelectric absorption due to metals present in the quasar X-ray
spectrum. The derived metallicity ranged from zero to solar
\citep{bechtold01,turnshek03}. Although, zero metallicity is unlikely
due to the observed metal lines in absorption
(Figure~\ref{fig:gasabs0312}), it can only be concluded that there is
no evidence for a relatively high metallicity DLA \citep{turnshek03}.
\citet{kanekar09} used STIS E230M quasar spectra to compute the
absorption metallicity using the {\ZnII} $\lambda2026$ and
$\lambda2062$ lines.  Zn abundance measurements give metallicity
estimates relatively free of depletion effects since Zn is relatively
undepleted on to dust grains.  \citet{kanekar09} compute a metallicity
of the $z_{abs}=0.313$ DLA to be [Zn/H]$=-0.90\pm0.11$ relative to
solar.


To compare the metallicity of this DLA to the general population of
DLAs, we use the work of \citet{kulkarni05} who computed the
N(\HI)-weighted mean [Zn/H] metallicity for 20 DLAs between $0.09 < z
< 1.37$. They derived a mean [Zn/H]$=-0.86\pm 0.11$ for their
maximum-limits sample, [Zn/H]$=-1.01\pm 0.14$ for their minimum-limits
sample. The maximum-limits sample treats the Zn limits as detections
and the minimum-limits sample treats the Zn limits as zeros. The
Q1127$-$145 $z=0.313$ DLA has roughly typical metallicity.


To summarize the metallicity comparison, we find that the gas in
absorption is relatively metal poor compared to the galaxies for which
we could compute metallicities.  We obtain roughly solar metallicity
for the galaxies and 1/10th solar for the absorption system.

For the $z=0.328$ galaxy pair, we compute the SFR for G5 using the
H$\alpha$ emission lines resulting in a
SFR(H$\alpha$)$=1.44$~M$_{\odot}$~yr$^{-1}$. The galaxy has a SFR much
lower than would be expected for strong outflows \citep{weiner09}.  We
are able to measure the metallicities for both galaxies (G3 and G5)
associated with the $z=0.328$ absorption.  G3 has a
[O/H]$=-0.10\pm0.09$ and G5 has a [O/H]$=0.09\pm 0.1$. That is, the
galaxies have similar metallicity which is approximately
solar. Unfortunately, since the absorption system is quite weak, we
cannot measure an absorption metallicity for this system for direct
comparison. Furthermore, from the {\it HST}/FOS G160L quasar spectrum
\citep[see][]{rao00} it is apparent that the $z=0.328$ Ly$\alpha$
absorption is embedded within the $z_{abs}=0.312710$ DLA Ly$\alpha$
absorption and cannot be deblended.

\begin{table}
\begin{center}
  \caption{Star formation rates and metallicities for galaxies
    associated with the $z_{abs}=0.328266$ and the $z_{abs}=0.312710$
    DLA. The galaxy star formation rates are not corrected for host
    galaxy dust extinction and the SFRs quoted are lower
    limits. However, even with applying such correction, the SFRs
    would not be strong enough to produce strong large-scale galactic
    winds. The errors in the metallicities include the statistical errors
    of the measurement of the emission line equivalent width
    measurements, fits to the continuum, and the errors in the
    empirical metallicity calibrators used.}
\vspace{-0.5em}
\label{tab:SFR}
{\footnotesize\begin{tabular}{lcccc}\hline
Gal-& SFR(\OII) & SFR(H$\alpha$) & 12$+$log(O/H) & [O/H]$^b$  \\
axy & M$_{\odot}$~yr$^{-1}$ & M$_{\odot}$~yr$^{-1}$ &    &   \\\hline
G3 & $\cdots$& $\cdots$  &  $8.64$$\pm$$ 0.05$ & $-0.10$$ \pm$$ 0.09 $ \\
G5 & $\cdots$&   1.44    &  $8.83$$\pm$$ 0.06$  & $\phantom{+}$$0.09$$\pm$$ 0.1$$\phantom{0}$ \\\hline
\multicolumn{4}{c}{$z_{abs}=0.313$ Group}\\\hline
G1 & $\cdots$& $\cdots$&  $\cdots$ & $\cdots$  \\ 
G2 &  0.44$\phantom{0}$     &   1.52    &  $8.74$$\pm$$0.05$  & $\phantom{+}$$0.00$$\pm$$0.09$ \\
G4 & 0.045 & $\cdots$&  $\cdots$ & $\cdots$  \\ 
G6 & $\cdots$& $\cdots$& $8.64$$\pm$$0.05$  &  $-0.10$$\pm$$0.09$ \\
G14 & $\cdots$& $\cdots$&    $<$8.43       &  $<-$0.31 \\\hline
\end{tabular}}
\end{center}
$^b$Here [X/Y]=log(X/Y)$-$log(X/Y)$_{\odot}$.
\end{table}

\section{Discussion}
\label{sec:dis}

\subsection{$z_{abs}=0.313$ Galaxies}

In most circumstances, DLAs are produced from quasar lines of sight
passing either through or near isolated galaxies
\citep[e.g.,][]{lacy03,moller02,rao03,chun06}.  However, in the case
of the $z=0.313$ DLA, the environment is more complex. Here we find a
group of galaxies associated with the DLA system containing at least
five members with a velocity dispersion $\sigma=115$~{\kms} offset
$80$~{\kms} blueward of the {\MgII} absorption redshift.  The DLA
associated with the $z=0.313$ group has $W_r(2796)=1.773$~\AA.  The
group as a luminosity range of $0.04\leq L_B\leq0.63 L_B^{\star}$ and
an impact parameter range of $17.4\leq D \leq 240.8$~kpc. The galaxy
redshift distribution is consistent with the {\MgII} absorption
velocity distribution, along with the other metals. Furthermore,
measured projected rotation curves of two of the galaxies (G2 and G4)
also cover the entire absorption velocity range.

From the derived galaxy emission line metallicities and the DLA
absorption metallicity, at first it appears unlikely that the
absorption is produced by metal-enriched winds or tidal debris from
these two galaxies.  Recent metallicity gradient measurements, derived
from local early type galaxies, have been shown to be quite shallow
and extend for several galaxy effective radii; \citet{spolaor10} find
an average stellar absorption metallicity gradient of $-0.22\pm0.14$
per effective radius for their sample. G2 has an effect radius of
$8.1\pm0.6$~kpc \citep{kacprzak07} which, with this metallicity
gradient, would imply that the metallicity at the quasar line of sight
would be [O/H]$=-1.2\pm0.7$.  Although the errors are large, this is
consistent with the absorption system metallicity observed. This
result is similar for G6. Thus, it is possible that the absorption is
produced by an extended disk that has a shallow metallicity
gradient. However, it has not been demonstrated that these metallicity
gradients can be smoothly extrapolated beyond a few disk effective
radii into the halo.


The only other direct comparison of galaxy emission line and quasar
absorption line metallicity was by \citet{bowen05} at redshift
$z=0.009$. The DLA has an impact parameter of 3.3~kpc from the
absorbing dwarf galaxy and the galaxy and the DLA have very similar
metallicities, perhaps implying a relatively flat radial abundance
gradient. This makes it difficult to determine if the gas originates
in the galaxy disk or outflows.  \citet{chen05} attempted to compare
host galaxy/absorber metallicities for three systems and found that
the galaxy metallicity derived from [O/H] greater than the absorption
metallicity derived from [Fe/H]. They proposed radial metallicity
gradients to explain their results. However, these results remain
uncertain because the absorption metallicities were derived using
iron, which has variable degree of dust depletion. Undepleted elements
such as Zn provide much more robust metallicity estimates which can be
more reliably compared to [O/H] metallicities derived for galaxies.

The outflow scenario is supported by \citet{bouche06} who found a
statistical anti-correlation between {\MgII} absorption-line
equivalent width and the mass of the halo hosting the absorbers by
cross-correlating absorbers with luminous red galaxies in the
SDSS. They claim that this is direct evidence that absorbers are not
virialized in gaseous halos of the galaxies. They suggest that the
strongest absorbers -- those with $W_r$(2796)~$\ga$~2~\AA, somewhat
stronger than the one associated with the $z=0.313$ group -- are
statistically more likely to trace super-winds.

Strong winds can be seen directly in absorption. For example, stacking
1400 DEEP2 galaxy spectra at $z$$\sim$1.4, \citet{weiner09} found
300--1000~km/s winds in {\MgII} and {\MgI} absorption in galaxies with
high SFRs and that both {\MgII} equivalent width and the outflow
velocities are correlated with galaxy SFRs.  At lower redshifts
($z$$\sim$0.6), \citet{tremonti07} reported {\MgII} absorption
blueshifted 500--2000 km/s relative to 14 post-starburst host
galaxies. Although some of these systems may not be DLAs,
\citet{rao06} showed that $\sim$35\% of (MgII-selected) absorbers with
\MgII$\lambda2796$ and \FeII$\lambda2600$ $ > 0.5$~{\AA} are DLAs.  We
find low SFRs in the two brightest galaxy members of the $z=0.313$
group (G2 and G4) and would not expect to see strong winds as
discussed above. However, we do not have SFR estimates for galaxy G1
which is at the closet projected distance to the quasar line of
sight. If this faint galaxy has/had high SFRs, then according to the
\citet{bouche06} results, this galaxy may have a high probability of
being a major contributor to the {\MgII} absorption. If the gas was
traveling at moderate wind speeds of 100~{\kms} then it would only
take $\sim 0.2$~Gyr to reach the quasar line of sight from G1. Though
this is a plausible argument, the tidal features seen for galaxies G2
and G4 suggest a different story.

This is not the first group discovered to be associated with {\MgII}
absorption. \citet{whiting06} detected a group of five galaxies
($z=0.66$) with a velocity dispersion of $\sigma=430$~{\kms}
associated with a {\MgII} absorption system with a velocity width of
250~\kms. Four of the five galaxies have impact parameters less than
100~kpc with the smallest at 51.2~kpc. They report that it is
difficult to associate a given galaxy to the absorption system and
that debris produced by interactions may be producing the absorption.
DLAs originating from tidal gas in galaxy groups are further supported
by \citet{nestor07} who found that very strong {\MgII} absorbers often
arise in fields with multiple galaxies in close proximity to the
quasar LOS. They do not have galaxy redshifts in the quasar fields,
yet they argue that the likely origin of the high equivalent width
{\MgII} absorption is kinematically disturbed gas around interacting
galaxies.  However, both studies used only ground based imaging and
were not able to directly study the morphologies of the group members.


For the $z_{abs}=0.313$ galaxy group, two of the brightest members
(G2, G4, and possibly G6) exhibit perturbed morphologies and several
extended optical streams. These streams extend for $\sim 25$~kpc and
may reflect the recent merger/interaction history of this galaxy
group; they may comprise tidal debris. The interactions producing the
tidal debris seen here may also be responsible for producing the
complex absorption system. The absorption could arise directly from
the tidal debris or from dwarf galaxies that form in these tidal tails
\citep{knierman03}.

Physical properties of the DLA absorbing gas are further constrained
by \citet{kanekar09} who derived a 21-cm gas covering fraction of 0.9
and a gas spin temperature of $T_s=820\pm145$~K. Spin temperatures of
$T_s\sim300$~K are typical for local spiral galaxies and the Milky
Way.  Higher spin temperatures are associated with smaller objects
such as dwarf galaxies, low surface brightness galaxies, or objects
with low metallicity/pressure that have a larger fraction of warm gas
and where physical conditions are not suitable for producing the cold
phase of HI \citep{wolfire95}.  However, the majority of DLAs
(including the $z_{abs}=0.313$ DLA) have far higher spin temperatures:
$T_s > 500$~K \citep{carilli96,kanekar09}.  It has been debated that
the high temperature estimates for DLAs may arise due to the
difference between radio and optical gas covering fractions
\citep{curran05} and/or wavelength dependent beam size or sightline
\citep{wolfe03}. Although these effects may play a role, a recently
reported correlation between DLA [Z/H] and $T_s$ may suggest that
there is no wavelength dependence on the observations
\citep{kanekar09}.  Given the large extended structure observed in the
radio for the Q1127$-$145 quasar, including the radio jet, it is
possible this system may suffer from such wavelength dependent effect,
thereby making it difficult to compare optical and 21-cm absorption
data. It has been mentioned in the literature that quasar Q1127$-$145
is a peculiar case, since it has a rather large 21-cm absorption
profile velocity width and a high spin temperature \citep{kanekar01b},
suggesting that this system is different from standard DLAs. Thus, it
is plausible that the both the optical and 21-cm absorption may arise
in structures such as tidal streams, infall and/or outflows.

The full velocity range of 350~{\kms} for the {\MgII} absorption
profiles (along with {\FeII}) remain difficult to reproduce in
kinematic models.  The cloud velocity distribution simulations of
\citet{prochaska97} suggest that DLA velocity profiles are driven by
rapidly-rotating thick disks. Also, 21-cm observations of low-mass
galaxies, such as the Large Magellanic Cloud, display lower velocity
widths than observed for typical DLAs \citep{prochaska02}. This
supports the idea that DLAs are massive rotating extended disks of
galaxies.  However, hydrodynamical simulations of \citet{haehnelt98}
showed that irregular protogalactic clumps can reproduce the DLA
absorption-line velocity width distribution equally well. They
conclude that the absorption velocity widths can be driven by a
variety of structures, which are a superposition of rotation, random
motions, infall, and merging. Additional 21-cm studies of
\citet{zwaan08} demonstrated that the DLA velocity widths do not
originate from rotating gas disks of galaxies similar to those seen in
the local universe. These results further support that DLAs are often
associated with tidal gas produced by galaxy interactions or
superwinds and outflows.

Given the data we have acquired, and the arguments that we have
presented, we favor the interpretation that the DLA absorption arises
from tidal debris produced by galaxy interactions, which are likely
more important in the $z=0.313$ group environment we have
identified. However, we cannot completely rule out other scenarios such
as outflows originating from the galaxy group members, faint
unidentified galaxies near the quasar LOS, or small satellite galaxies
in front of the quasar LOS.
 
\subsection{$z_{abs}=0.328$ Galaxies}

The $z=0.328$ pair of galaxies is associated with a weak {\MgII}
absorption system. Both galaxies, G3 and G5, are within the fiducial
{\MgII} halo size of $\sim$100~kpc. G5 is roughly 2.5 times more
luminous than G3.  Both galaxies have velocities that are consistent
with the absorption velocities (see
Figure~\ref{fig:gasabs0328}), which makes it difficult to
associate one particular galaxy to the absorption system. The SFR of
G5 is typical of a normal spiral galaxy and would not be expected to
have strong outflows \citep[e.g.,][]{heckman02,heckman03,weiner09}.

Both galaxies have similar metallicities which are roughly solar. Even
if we had the metallicity of the absorption system we would not be
able to identify the host galaxy.  For reference, \citet{narayanan08}
analyzed 100 weak {\MgII} absorbers and found, using ionization
modeling, that the metallicity in a significant fraction of systems
are constrained to values of solar or higher. If this was true for
this particular case, both galaxies would be in agreement with the
absorption metallicity.  There is no clear evidence of strong
disruptions in the morphology of either galaxy, indicating no recent
merger or interaction activity.  Given that the absorption system is
very weak, it could arise in a wide array of structures associated
with the environment of the pair of galaxies.

\section{Conclusions}
\label{sec:conclusion}

We have performed a spectroscopic galaxy survey to limiting magnitude
of $m_{F814W}\leq 20.3$ ($L_B>0.15L_*$ at $z=0.3$) within
100$\times$100$''$ of the quasar Q1127$-$145.  This field has a large
number of bright galaxies near the quasar line of sight and has three
{\MgII} absorption systems detected in the quasar spectrum, including
one DLA.  Here we have obtained spectroscopic redshifts for eight
galaxies in this field, adding to the four previously identified
\citep{bb91,gb97,kacprzak10}.

Our main results can be summarized as follows:

\begin{enumerate}

\item We have identified two galaxies (G6 and G14) associated with the
  DLA at $z=0.313$, which, in addition to the three known galaxies,
  form a group of at least five galaxies.  The group has a luminosity
  range of $0.04\leq L_B\leq0.63 L_B^{\star}$ and an impact parameter
  range of $17.4\leq D \leq 240.8$~kpc.  The group velocity dispersion
  is $\sigma=115$~{\kms} having a full velocity range of $\sim
  350$~\kms.  The group redshift is offset $80$~{\kms} blueward of the
  {\MgII} absorption redshift. The galaxy redshift distribution spans
  the entire range of the absorption velocities. Furthermore, the
  rotation curves of G2 and G4 alone cover the entire range of
  absorption velocities.

  Star formation rates of two of the brightest galaxy members are too
  low to drive strong winds, reducing the likelihood that winds are
  responsible for the absorbing gas. Metal enriched winds are also
  unlikely since the DLA metallicity is 1/10th solar, whereas three of
  the five galaxies have metallicities range between less than 1/2
  solar to solar. Although stellar metallicity gradients in the
  literature are consistent with our findings, it is has yet to be
  demonstrated that these gradients can be extrapolated to 50~kpc. The
  favored scenario for the origin of the absorption is from tidal
  debris. The deep WFPC--2 F814W imaging shows the perturbed
  morphologies for three galaxies and optical tidal tails extending
  $\sim 25$~kpc away from the disks. These features suggest
  merger/harassment events, consistent with the more frequent galaxy
  harassment/merging expected in the group environment we have
  identified.

\item We have identified a galaxy (G3), in addition to previously
identified G5 \citep{kacprzak10}, associated with the $z=0.328$ weak
{\MgII} absorption system, $W_r(2796)=0.029$~\AA.  There is no evidence of recent
interactions since both galaxies have unperturbed morphologies
 and they are separated by 140~kpc.  Even
armed with the star-formation rate and rotation velocities of G5 and
the metallicities of both galaxies, it remains difficult to determine
which galaxy hosts the absorber.  We can only conclude that this
weak absorption system can arise in a variety of cosmic structures in
either or both halos of the galaxy pair.

\item We have identified five galaxies (G7, G9, G11, G12, and G15)
with $0.21\leq z\leq0.33$ that are not associated with any detectable
{\MgII} absorption (3$\sigma$ detection limits of $4.8-5.7$~m{\AA}).
These galaxies appear to be normal star-forming spiral disks. All
non-absorbing galaxies have impact parameters $D>118$~kpc. This is
consistent with previous results on {\MgII} halo sizes, which suggest
we should not expect to detect absorption beyond impact parameters of
$\sim 120$~kpc.

\end{enumerate}

The DLA-galaxy group at $z=0.313$ is quite different from the standard
examples in literature of DLA-plus-(apparently) isolated galaxy
\citep[e.g.,][]{lacy03,moller02,rao03,chun06}.  The group of galaxies
associated with the $z=0.313$ DLA suggests that interactions, which
are common in groups of galaxies, might be responsible for at least
some DLA absorption systems as well. This may explain why searches for
host galaxies of DLAs and strong {\MgII} systems have a low success
rate of 30--40\% using small field of view IFUs
\citep[e.g.,][]{bouche07}.  It is likely that we need to survey
further out from the quasar line of sight if there are many other
cases where tidal debris produces the absorption.  It is also
interesting to note that if this galaxy group was at a slightly higher
redshift, we would not be able to detect the 0.04$L_B^{\star}$ galaxy
that is closest to the quasar line of sight, which could even be the
DLA host. Given the low redshift of the DLA and even using the deep
{\it HST} imaging, star formation rates, and metallicities, it is
difficult to understand this complex system and determine the origins
of the absorbing gas. We emphasize that we should take caution in
concluding the origins of absorbing gas drawn from studies of
individual DLAs at higher redshifts.


\section*{Acknowledgments}

We thank Frank Briggs for his useful comments and for carefully
reading this paper.  We thank Greg Wirth for his help and advice with
ESI/Keck.  MTM thanks the Australian Research Council for a QEII
Research Fellowship (DP0877998). CWC was supported by the National
Science Foundation under Grant Number AST-0708210. This work is based
on observations obtained with the Apache Point Observatory 3.5-meter
telescope, which is owned and operated by the Astrophysical Research
Consortium.  Observations were also made with the NASA/ESA {\it Hubble
Space Telescope}, or obtained from the data archive at the Space
Telescope Institute.  Other observations were made with the ESO Very
Large Telescope at the Paranal Observatories. Based on observations
made with the NASA/ESA Hubble Space Telescope, and obtained from the
Hubble Legacy Archive, which is a collaboration between the Space
Telescope Science Institute (STScI/NASA), the Space Telescope European
Coordinating Facility (ST-ECF/ESA) and the Canadian Astronomy Data
Centre (CADC/NRC/CSA). Some of the data presented herein were obtained
at the W.M. Keck Observatory, which is operated as a scientific
partnership among the California Institute of Technology, the
University of California and the National Aeronautics and Space
Administration. The Observatory was made possible by the generous
financial support of the W.M. Keck Foundation.


\end{document}